\documentclass{iauc}
\usepackage{graphicx}

\title[Age and Metallicity in Old Stellar Populations] %% give here short title %% 
{Age and Metallicity Estimations in Old Stellar Populations from Str\"omgren Photometry}

\author[Rakos, Schombert \& Odell]   %% give here short author list %%
{K. Rakos$^1$, J. Schombert$^2$  \& A. Odell$^3$}

\affiliation{$^1$Institute for Astronomy, University of Wien
\\[\affilskip]
$^2$Department of Physics, University of Oregon
\\[\affilskip]
$^3$Department of Physics, Northern Arizona University
}

\pubyear{2005}
\volume{198} %% insert here IAU Symposium No.
\pagerange{1--8}
\date{28 Apr 2005}
\jname{Proceedings Title IAU Colloquium }
\editors{A.C. Editor, B.D. Editor and C.E. Editor, eds.}
\begin{document}

\maketitle

\begin{abstract}

We present a new technique to determine age and metallicity of old stellar populations
(globular clusters and elliptical galaxies) using an iterative principal component analysis on
narrow band (Str\"omgren) colors.  Our technique is capable of reproducing globular cluster
[Fe/H] values to 0.02 dex and CMD ages to 1.0 Gyrs.  We also present preliminary results on the
application of our technique to a sample of high mass, field ellipticals and low mass, cluster
dwarf ellipticals.  We confirm the results of earlier studies which find that globular clusters
increase in metallicity with age and that age and metallicity increase with galaxy mass.
However, we find that dwarf ellipticals deviate from the elliptical sequence by having little
to no correlation between age and metallicity.


\end{abstract}

\firstsection % if your document starts with a section,

\section{Introduction}

The two primary processes that determine the characteristics of a stellar population are its
star formation history and its chemical evolution.  For an actively star forming system, such
as the disk of our Galaxy, these two process are intertwined and will display a feedback loop as
star formation continues and, thus, understanding an active system requires detailed HR
diagrams and individual stellar spectroscopy.  However, a simple stellar population (SSP), one
formed in a single event from a single cloud of gas (e.g. a globular cluster), will have a
fixed metallicity and age that may be derived from the color-magnitude diagram (CMD).  A burst
stellar population, i.e. one derived from a extended star formation event, will be composed of
a combination of SSP's and the evolutionary processes are reflected into the population's age and
metallicity by the luminosity weighted mean of the various SSP's. It is possible to
characterize a burst population if the duration of the burst is short and distribution of
metallicities is uniform (Renzini \& Buzzoni 1986).

Early studies of composite stellar populations focused on broadband colors of spiral bulges and
elliptical galaxies (Sandage \& Visvanathan 1978, Tinsley 1980, Frogel 1985) and these datasets
supported the hypothesis that red galaxies are composed, primarily, of old, metal-rich stellar
populations under the burst hypothesis.  Unfortunately, it was quickly realized that detailed
interpretation of broadband colors with respect to age and metallicity are complicated by
several factors.  Foremost was the assumption that old stellar systems are composed of a
uniform population in age and metallicity.  It was soon demonstrated by population synthesis
techniques (O'Connell 1980) that a young population quickly reddens to similar integrated
colors as an old population and that the change in color is abrupt even while the differences
in age may be quite large (greater than 5 Gyrs).  In addition, it was identified through the
use of stellar population models that slight changes in age and metallicity operate in the same
direction of spectroevolutionary parameter space (Worthey 1994).  This coupling of age and
metallicity (known as age-metallicity degeneracy, Worthey 1999) is due to competing
contributions from main sequence turn-off stars (sensitive to age) and red giant branch (RGB)
stars (sensitive to metallicity) near 5000\AA.  Filters that bracket this region of a galaxy's
spectrum will require increasingly accurate values for metallicity to determine a unique age
and vice-versa.

To avoid the age-metallicity degeneracy problems, a majority of recent stellar population
studies have focused on the determination of age and metallicity through the use of various
spectral signatures, such as H$\beta$ for age (Kuntscher 2000, Trager \etal 2001).  This
approach provides a finer comparison to stellar population models, but requires assumptions
about the relationship between metallicity indicators (e.g.  Mg$_2$) and the [Fe/H] value of
the population as reflected into the behavior of the red giant branch.  In other words,
spectral lines provide the value of that element's abundance, but what is really required is
the temperature of the RGB which is a function of the total metallicity, $Z$.  Varying ratios
of individual elements to $Z$ complicates the interpretation of line studies (Ferreras, Charlot
\& Silk 1999).  In addition, these techniques have limitations due to the required high S/N for
the data that make them problematic for the study of high redshift systems.

An alternative approach to spectral line studies is to examine the shape of specific portions
of a spectral energy distribution (SED) using narrow band filters centered on regions sensitive
to the mean color of the RGB (metallicity) and the main sequence turnoff point (mean age)
without the overlap that degrades broadband colors.  The type of galaxy examined, for example
stellar systems with ongoing star formation where a mix of different age populations may be
present, will still limit this technique.  However, for systems that have exhausted their gas
supply many Gyrs ago (i.e. old and quiescent), it may be possible to resolve the underlying
population with some simple assumptions on their star formation history and subsequent chemical
evolution.  Thus, we have the expectation, guided by the results of evolution models, that
objects composed of SSPs or a composite of SSPs (e.g. ellipticals) present special
circumstances where the age-metallicity degeneracy can be resolved and allow the study of the
evolution of stellar populations.

\begin{figure}
\includegraphics[width=9.5cm,angle=-90]{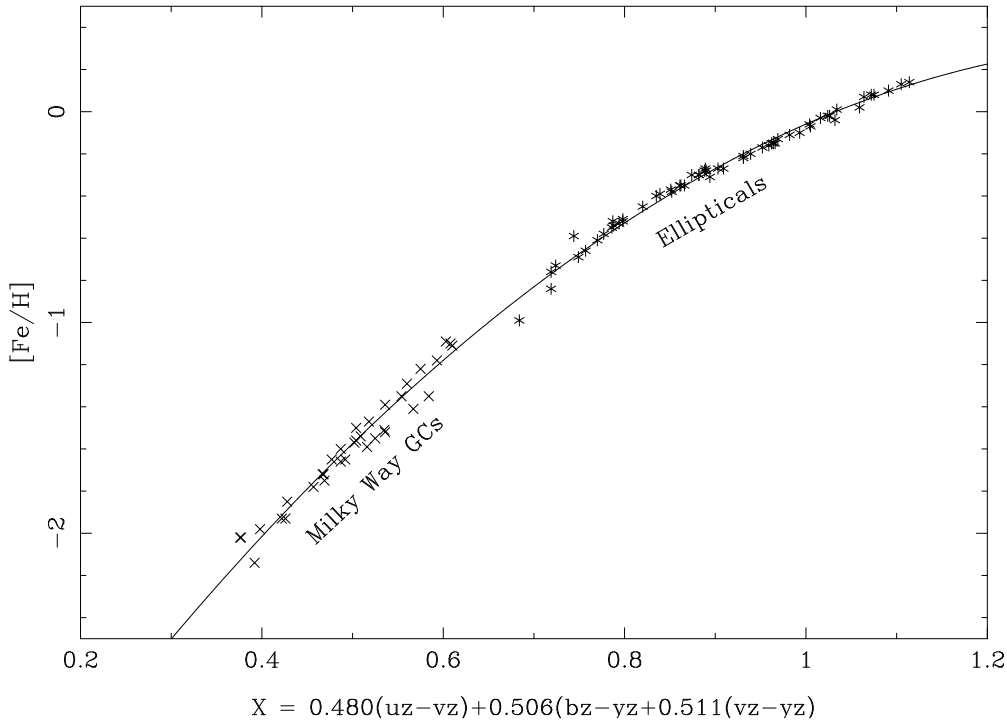}
\caption{The mean metallicity of the globular cluster and elliptical samples as determined by
the iterative technique versus the multi-color term that incorporates all of the PC terms.
This figure demonstrates how the use of the full parameter space reduces the scatter in [Fe/H].
Also shown are the iterative results for a sample of dwarf and giant ellipticals in the Coma
cluster with full $uvby$ photometry.  The relationship become less linear for galaxies due to
the fact that they are composed of a integrated population of SSP's.  The low scatter indicates
that metallicity is the primary driver of galaxy color (see also Smolcic \etal 2004).}
\end{figure}

In a series of earlier papers, we have examined the modified Str\"omgren $uz,vz,bz,yz$ colors
of globular clusters and used a combination of their colors and SED models to derive the mean
age and metallicity of dwarf, bright and field ellipticals (Rakos \etal 2001, Odell, Schombert
\& Rakos 2002, Rakos \& Schombert 2004).  While spectroscopic data is superior for age and
metallicity estimations in high S/N datasets, our goal has been to develop a photometric system
that can be used for galaxies of low surface brightness and/or high redshift, where
spectroscopy is impractical or impossible.  In addition, we believe it is important to compare
continuum based age and metallicity estimates to line indice estimates to expand the range of
testable observables for our spectroevolutionary models.

\begin{figure}
\includegraphics[width=9.5cm,angle=-90]{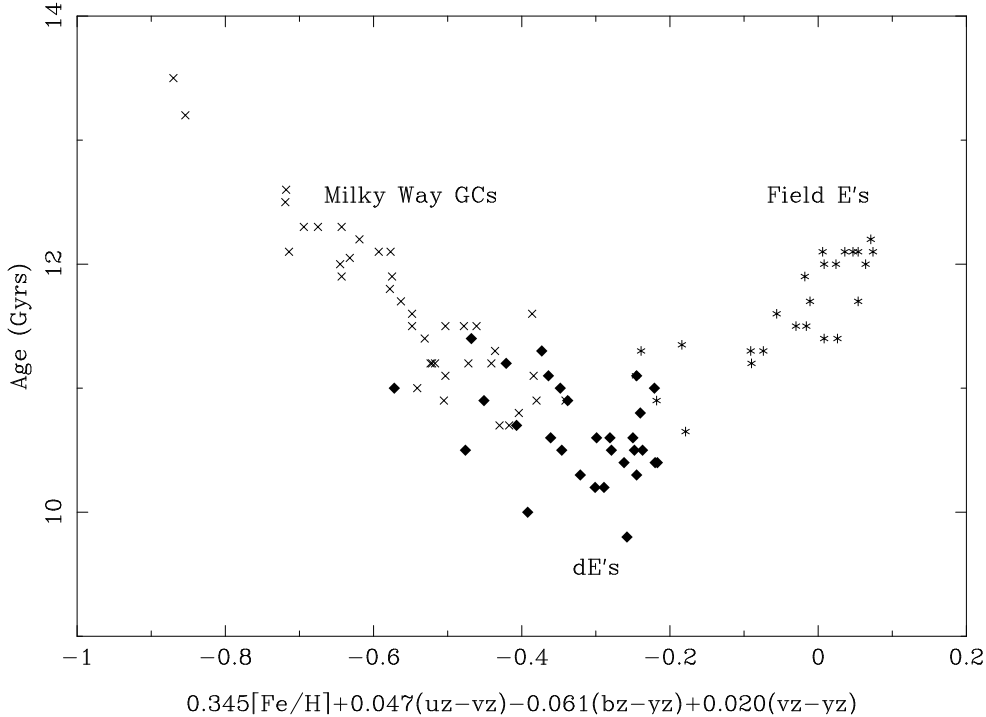}
\caption{Mean stellar age, as determined by an iterative PC fit, versus an "ageless" color
term.  While any portion of this diagram can be occupied by a range of age and metallicities,
in fact, old, metal-poor objects (globulars) and old, metal-rich objects (ellipticals) follow
two separate linear correlations.  Dwarf ellipticals bridge region between globulars and
ellipticals as transition objects with a smaller range of metallicity in their underlying
stellar populations.
}
\end{figure}

Our past technique has been to relate the $vz-yz$ color index to mean metallicity, since the
$vz$ filter is centered on the absorption line region near 4100\AA\, and the $bz-yz$ color
index, whose filters are centered on continuum regions of the spectral energy curve, measures
the mean stellar age.  These results, as guided by comparison to multi-metallicity SED models,
was crude since metallicity changes will move the effective temperature of the RGB and, thus,
the continuum $bz-yz$ colors.  In our more recent work, we have included photometry through the
$uz$ filter which provides an additional handle on age and metallicity effects to the other two
color indices.  In addition, we have applied a principal component (PC) analysis on the
multi-color data (Steindling, Brosch \& Rakos 2001) that more fully isolates metallicity from
age effects and the changes due to recent star formation.  In the end, we believe we have
isolated a reliable technique for estimating age and metallicity for non-starforming objects
such as ellipticals and S0's.  This paper will describe some of our preliminary results in
comparing the star formation history of globular clusters, field ellipticals and dwarf
ellipticals.

\section{Age and Metallicity Calibration}

The technique we have developed to apply to our narrow band filter system is full described in
Rakos \& Schombert (2005).  In brief, our method hinges using principal component (PC) analysis
on the multi-color phase space generated by a unique set of SED models (Schulz \etal 2002).  A
PC analysis in any $n$-dimensional space, calculates the axis along which the model points
present the largest, most significant scatter.  This is called the first principal component
(PC1). PC analysis then proceeds to calculate PC2, the axis of the second most significant spread
in the remaining $n-1$ dimensional space orthogonal to the first PC, and so on.  While
scientific data in the astronomy are usually neither linear nor orthogonal at the same time.
Often, for small regions of data, linearity and orthogonality can be reached through the use of
PC analysis.  Close inspection of Schulz \etal theoretical models has shown acceptable
linearity for ages larger than 3 Gyrs over the full range of model metallicities.  In this
restricted region, it is possible to apply PC analysis to; 1) separate the age and the
metallicity of a stellar population, 2) select the most correlated variables and 3) determine
linear combinations of variables for extrapolation.  

\begin{figure}
\includegraphics[width=9.5cm,angle=-90]{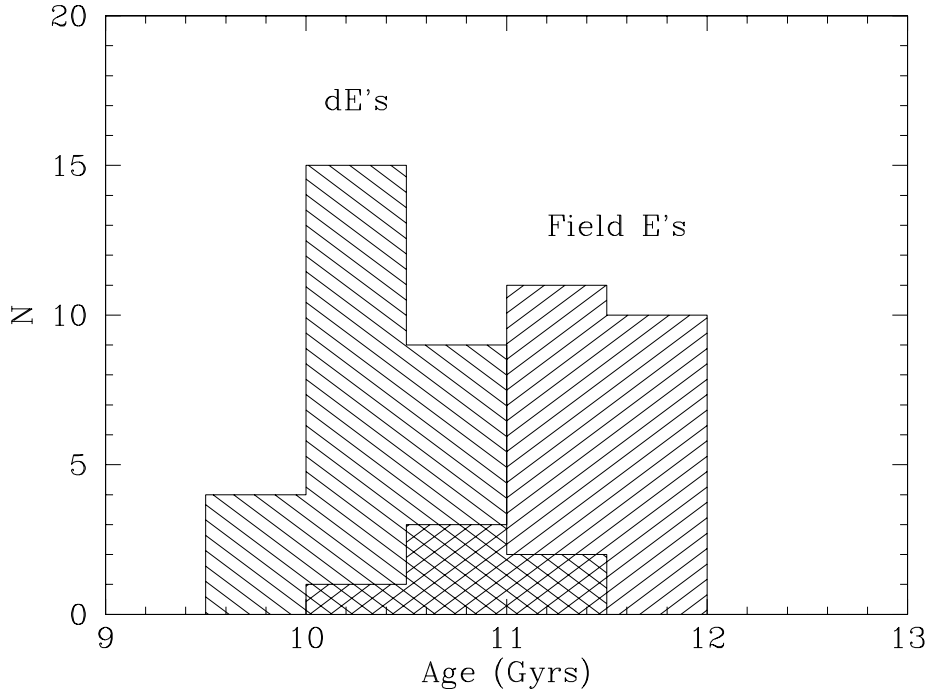}
\caption{The distribution of mean stellar age for dwarf ellipticals (dEN's from Fornax) and
field ellipticals (a bright, nearby sample).  The mean dE age is 1.0 Gyrs younger than the mean
for field ellipticals.  This difference can reflect an earlier star formation epoch for field
ellipticals or a prolonged initial stage of star formation for dE's.
}
\end{figure}

The advantage of the smooth behavior for age and metallicity in the PC plane is that if one has
all three narrow band colors ($uz-vz$, $vz-yz$ and $bz-yz$) then knowledge of the correct PC1
and PC2 values allows for the unique determination of age and metallicity from the PC
equations.  If the PC values are unknown, then an iterative search scheme could select a range
of values for mean age and [Fe/H], determine PC1 and PC2 from the model fits, then compare how
well those values compare with the results from the PC equations.  To test this iterative
procedure, we ran a sample of 40 globular clusters with known ages (CMD fitting) and
metallicities (spectral) where we have obtained high precision $uvby$ photometry.  The resulting
fits re-captured the correct ages and [Fe/H] values to an accuracy of 0.5 Gyrs in age and 0.05
dex in [Fe/H].  Additional analysis with the iterative solutions finds that for both globular
clusters (simple SSP's) and ellipticals (composite SSP's), it is possible to find a highly
accuracy metallicity simply from a linear combination of narrow colors, by themselves (see
Figure 1).

This technique differs from previous line indice work in several regards.  First, the narrow
band filters used in our study are specifically selected to avoid major line regions of the
spectrum.  Thus, they provide a measure of the integrated continuum of a stellar population and
the value for Fe/H determined by this method will be based on stellar atmospheric temperatures
rather than individual line depth.  While this may be less accurate than direct determination
of the strength of Fe in a spectrum, it has the advantage of being the quantity you wish to
know about a stellar population (its mean metallicity $Z$ as it effects the colors of the
stars) and does not suffer from calibration shifts when going from one line feature to Fe/H
(e.g.  changing ratios of Mg/Fe in ellipticals).  The metallicity values produced by this
method will also be luminosity weighted means for the entire stellar population.  This is in
contrast to line indice studies which are surface brightness selected and will be strongly
biased by central core light.

\section{Application to Globular Clusters and Ellipticals}

We have three new sources of photometry in which to test our technique, Milky Way globular
clusters (described in the last section), 25 nearby field ellipticals and a sample of 40 dwarf
ellipticals in Fornax.  From our previous work with ellipticals, we know that their narrow band
colors cannot be described by a single metallicity/age SSP (Rakos \etal 2001).  However, we
found that models which combine a range models (roughly gaussian) of varying metallicities in a
linear fashion accurately reproduce the color of ellipticals.  Each set of models can be
parameterized by a luminosity weighted mean metallicity and that decreasing metallicity with
mass (the color-magnitude) effect reflects a decreasing spread or range in underlying
population metallicities.  Thus, at high luminosities, ellipticals are a composite of many
stellar populations, but at low luminosities (mass), the colors of ellipticals approach the
colors of SSP's.

\begin{figure}
\includegraphics[width=9.5cm,angle=-90]{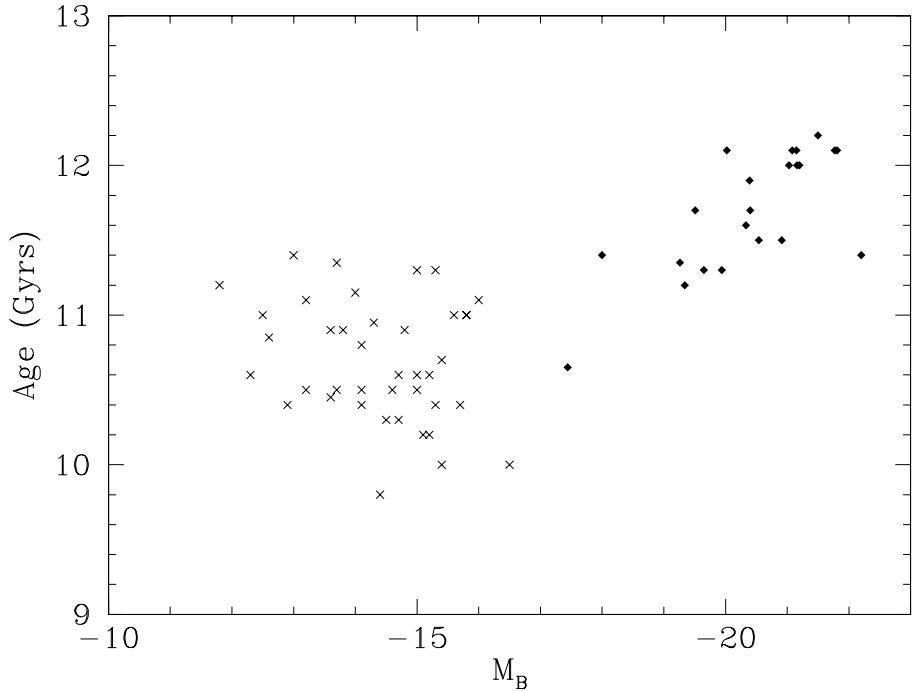}
\caption{Mean stellar age versus galaxy blue luminosity (mass).  There is a clear correlation
for field ellipticals of decreasing age with decreasing mass, but dwarf ellipticals little to
no correlation with age (a slight tendency to find increasing age with decreasing mass).
Combined with the color-magnitude (mass-metallicity) relation for bright ellipticals, there is
no anti-correlation between age and metallicity as predicted by hierarchical models of galaxy
formation.
}
\end{figure}

We, therefore, have the expectation that dwarf ellipticals would bridge the gap between
ellipticals and globular clusters in the PC space defined by our three narrow band colors, age
and mean metallicity.  This can be seen in Figure 2, a plot of galaxy age, as determined by our
iterative PC technique and an "ageless" PC term given by the linear combination of metallicity
and the three narrow band colors (note that the value of [Fe/H] is the one determined by the PC
fits, although a value of [Fe/H] from relationship in Figure 1 would have produced identical
results).

There are several important points to note about Figure 2. The first is that any region of this
diagram can be occupied by various combinations of high or low metallicity versus old or young
mean stellar age.  Our samples divided into old, low metallicity objects (globular clusters)
and old, high metallicity objects (high mass ellipticals).  The addition of the dwarf
elliptical data joins the globular and elliptical sequences at low metallicity, younger age.
As found in our previous work on dwarf ellipticals (Rakos \& Schombert 2004).  Nucleated dwarfs
(dEN's) track the globular sequence colors to higher metallicities and, as in Figure 2, younger
age.

A direct comparison of the ages of field and dwarf elliptical is found in Figure 3, a histogram
of their PC calculated ages.  While the initial interpretation of this Figure is that dwarf
ellipticals are younger than field ellipticals (hierarchical models of galaxy formation predict
that low mass galaxies are older than high mass systems, but also predict that field galaxies
are older than cluster galaxies), "age" from these calculations refers to mean stellar age.  A
"younger" mean age can be obtained through a later epoch of initial star formation (not
necessary when the dark+baryonic matter lump formed) or through a prolonged era of initial star
formation.  And, in the case of dwarf galaxies, later weak bursts of star formation would also
reflect into a younger mean age.

While as a group the dE's in Fornax have lower mean stellar ages than field ellipticals, this
trend is not as linear with luminosity (mass) as one would expect.  Figure 4 displays the run
of age versus absolute blue luminosity (M$_B$).  The trend for field ellipticals is a clear one
of decreasing age with decreasing mass.  This parallels the well-known mass-metallicity
relation such that older galaxies have higher metallicities (as predicted by closed box models
of chemical evolution, Matteucci 2003).  However, the dwarfs display little correlation with
mass (similar to the breakdown of the narrow band color-magnitude relation for dwarf galaxies,
Odell, Schombert \& Rakos 2002) which may signal that higher mass ellipticals are constructed
in a uniform fashion, but that low mass dwarf ellipticals have episodic past histories of star
formation.

\begin{figure}
\includegraphics[width=9.5cm,angle=-90]{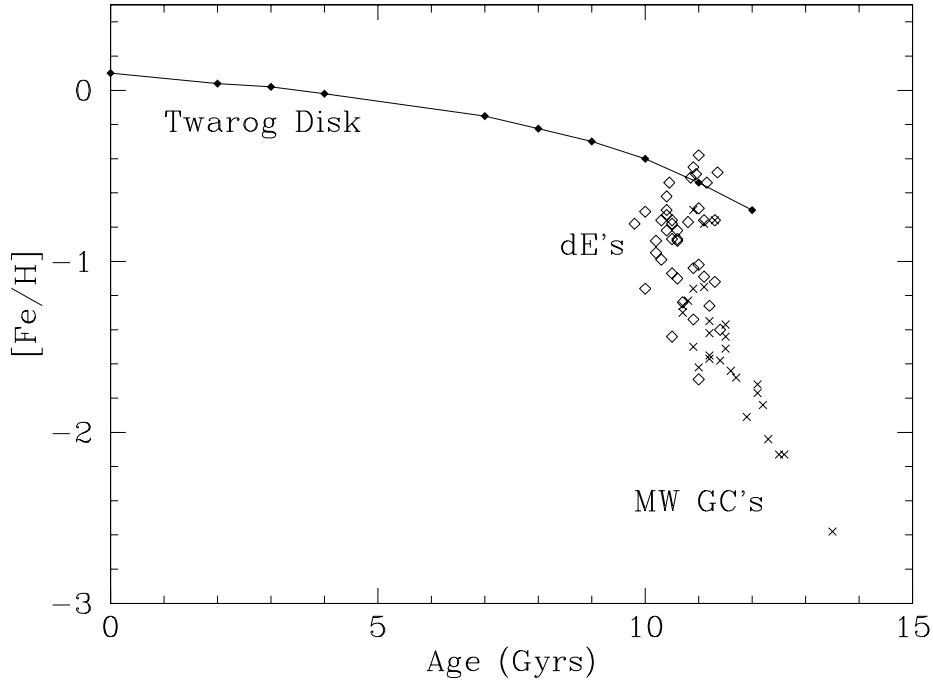}
\caption{The age-metallicity diagram for Milky Way globular clusters and Fornax dwarf
ellipticals.  The age-metallicity relation for solar neighborhood F dwarfs (Twarog 1980) is
also shown.
}
\end{figure}

Given the trend  $\alpha$/Fe ratios in ellipticals (Thomas \etal 2005), which measures the
duration of the initial star formation burst, such that high mass ellipticals have shorter
durations, part of the trend in Figure 4 is from the prolonged star formation for low mass
ellipticals.  Between a luminosity of $-$17 and $-$22, Thomas \etal estimate a decrease in
duration of approximately 0.75 Gyrs, which is close to the age of 1.5 Gyrs for the same mass
range in Figure 4.  Thus, bright ellipticals may be coeval, but with a slowly varying duration
of initial star formation that reflects into younger mean age.  Testing this hypothesis to a
larger sample of cluster ellipticals will be complicated by recent mergers of star-forming
galaxies which artificially lower the mean stellar age (Trager \etal 2001)

\section{Chemical Evolution of Old Stellar Populations}

Armed with a handful of age and metallicity values for Milky Way globular clusters and dwarf
ellipticals, we can investigate the earliest epoch of the construction of the age-metallicity
relation (AMR, Twarog 1980).  Figure 5 displays the age and [Fe/H] values for the low
metallicity region of the PC phase space.  Also shown in Figure 5 is the AMR for the solar neighborhood as
determined by Str\"omgren plus H$\beta$ photometry of F dwarfs.  The steeper slope to the
galaxy data reflects the expected rapid enrichment process predicted by closed box models of
chemical evolution (Bond 1981).

The era where the dE data ends corresponds roughly to a [Fe/H] $\sim$ $-$0.6 which is the
transition point between the Galactic disk, where the closed models for chemical evolution are
inadequate, and the halo component (Gilmore \& Reid 1983).  While the dwarf elliptical data
lacks the clear correlation between age and metallicity seen in the globular cluster data, it
is obvious that the low metallicity dE's follow the globular clusters and that above [Fe/H] =
$-$0.8 the dE's decouple from a simple model of chemical evolution (i.e. a closed box with
instantaneous recycling).  This is consistent with our earlier interpretation that the lower
metallicity dwarfs have a very narrow range of internal metallicities and that their underlying
stellar populations, to first order, mimic the colors of an SSP, such as a globular cluster.

\begin{acknowledgments}

We wish to thank all the observatories (NOAO, Steward, ESO) which have granted us time for this project.  And we
are grateful to the organizers of this Colloquium for a chance to present our newest results.

\end{acknowledgments}

\end{document}